%%%%%%%%%%%%%%%%%%%%%%%%%%%%%%%%%%%%%%%%
%%%%  ZA SPIG 1998  %%%%%%%%%%%%%%%%%%%%%
%%%% verzija od nedelje 24.5.98 
%%%% SA  SKRACCIVANJEM
%%%%%%%%%%%%%%%%%%%%%%%%%%%%%%%%%%%%%%%%%
\input psfig
\magnification=\magstep1

\font\naslov=cmbx10 scaled \magstep1
\font\adresa=cmbxti10

\def\frac#1#2{{\begingroup#1\endgroup\over#2}}

\centerline{\naslov ON THE IONIZING SOURCES IN SPIRAL GALAXIES:}

\centerline{\naslov I. FROM CENTRAL REGIONS TO THICK DISK}

\centerline{\naslov  AND DISK-HALO CONNECTION}

\medskip

\centerline{\bf  S. Samurovi{\'c}$^{1}$,  M.  M. {\'C}irkovi{\'c}$^{2,3}$, 
V. Milo{\v s}evi{\'c}--Zdjelar$^{1}$
 }

\centerline{\bf and  J. Petrovi{\'c}$^4$}

\medskip

\centerline{\adresa $^1$ Public Obs., Gornji Grad 16, Kalemegdan, 11000
Belgrade, Yugoslavia}

%\centerline{\it $^*$E-mail srdjanss@afrodita.rcub.bg.ac.yu}

%\centerline{\it $^2$ Public Obs., Gornji Grad 16, Kalemegdan, 11000
%Belgrade, Yugoslavia}

%\centerline{\it $^{**}$ E-mail vesnamz@afrodita.rcub.bg.ac.yu}

%\medskip

\centerline{\adresa $^2$  SUNY at Stony Brook,  Stony Brook, NY 11794-3800, USA}

\centerline{\adresa $^3$ Astronomical Obs., Volgina 7, 11000 Belgrade,
Yugoslavia}

%\centerline{\it E-mail  cirkovic@sbast3.ess.sunysb.edu}

%\medskip

\centerline{\adresa $^4$ Dept. of Astronomy, Stud. trg, 11000 Belgrade, Yugoslavia}

\bigskip
\noindent{\bf 1. INTRODUCTION}
\medskip

\noindent The problem of the ionization in the spiral galaxies, including our 
own, the Milky Way is not solved at the moment (for survey see e.g. [1]) 
although there are some indications that could lead towards solution [2, 3, 
4].

We assume  the Sofue's model of the rotation curves of spiral 
galaxies [5] based upon the observation of CO line emission and 
combined with HI and optical observations. This model  has four 
mass components: nuclear mass component,  a central bulge,
 a disk,  and a massive halo. One can add the fifth component for the 
Milky Way [5] at the nucleus e.g. [6,7].
As a characteristic  of the ionization we study the number density of free 
electrons in the interstellar medium throughout the Galaxy, $n_e$, using the 
Taylor and Cordes model (TC93) [8] as well as  its 
updated version (LC98) [9].

\bigskip \noindent{\bf 2. THE CASE OF THE MILKY WAY} \medskip

\noindent We examine our Galaxy -- Milky Way, in many respects a typical 
spiral galaxy, in more detail. In the 
textbooks one usually encounters the fact that in the disk of a spiral there
are  H~II regions that are ionized by the radiation from young and massive
 stars,  associations of OB stars [10].
 Although the average 
properties of the ionized gas are rather well known the source of this 
ionization is still the subject of the discussion of the astronomers, 
astrophysicists,   and even cosmologists. Following Sciama [1] one can 
state the following important problems. 

\item {$\bullet$} The scale height (defined as the column density on one side 
of the galactic plane divided by the volume density in the plane) of free 
electron component of the interstellar matter (ISM) is $h_e=\left ( 
{N\over \langle n_e \rangle_0}\right )
\approx  668$ 
pc, where $N$ is the the total column density ($N\sim 6.8 \times 10^{19}$ 
${\rm cm}^{-2}$ [11]) and $\langle n_e \rangle_0=0.033$ ${\rm 
cm^{-3}}$ is the mean electron density at the midplane.  However, the sources such 
as O stars or supernovae have a much smaller scale height ($h_e\sim 100$ pc).

\item {$\bullet$} As pointed out by Reynolds  [12] the nature and origin of 
the ionized gas about the galactic disk is not well understood.  The power 
that is required to maintain the diffuse, ionized gas ranges from minimum of 
$5.3\times 10^{-5}$ ergs s$^{-1}$ cm$^{-2}$ (if all the gas has $T\sim 10^4$ 
K) to  $6.6\times 10^{-3}$  ergs s$^{-1}$ cm$^{-2}$ ($T\sim 10^5$ K). As 
stressed by  [1] among all known sources of power 
only Lyc (Lyman continuum) photons from O stars and energy from supernovae 
have or exceed the required power. 

\item {$\bullet$} Therefore, since the interstellar medium is highly opaque to 
hydrogen-ionizing radiation, the following question remains: how can this 
radiation travel hundreds of parsecs from the parent O stars in order to 
produce the diffuse ionized gas? (e.g. [13]). Careful study of 
the interstellar hydrogen towards different pulsars suggests that column 
densities of free electrons along two line segments cannot be 
 accounted for by H~II regions around B stars or hot white dwarf stars. 
Reynolds [12]  gives three possibilities for the explanation of the 
existence of the ionized gas: (i) morphology for the interstellar H~I is very 
different from that usually depicted, (ii) Lyc luminosities for early B or hot 
white dwarf stars are more than an order of magnitude larger that the 
currently accepted values and (iii) there exists an additional, yet 
unrecognized source of ionization within the Galactic disk.

\item {$\bullet$} The mean electron density in opaque intercloud regions 
within a few hundred parsecs of the Sun as function of a Galactocentric 
radius is approximately constant~[8]. 

TC93 modeled the electron density of a certain Galactic location as the
 following sum (the fifth component  was added in LC98):
\vskip-0.5cm
$$n_e(x,y,z)=n_1g_1(r){\rm sech}^2(z/h_1)+
n_2g_2(r){\rm sech}^2(z/h_2)+$$
\vskip-0.7cm
$$n_a{\rm sech}^2(z/h_a)\sum _{j=1}^4 f_jg_a(r,s_j)+
n_Gg_G(u)+
n_{\rm GC}g_{\rm GC}(r)h_{\rm GC}(z)\eqno(1)$$
\vskip-0.35cm
\noindent The detailed description of each component is given in TC93 and LC98 and 
 because of space limitation we just give  few remarks: $r$ is the
 Galactocentric
distance projected onto the plane and is equal $r=(x^2+y^2)^{1/2}$, the 
sum goes over four spiral arms, $n_i\; ,\,  i=1\dots 5$ denotes the density
in different regions, $f_i\; ,\, i=1\dots 4$ are scale factors,
 $g_i\;, \, i=1\dots 5$ are functions of position, $h_i,\; i=1\dots 4$ are
scale heights and $z$ is the height above the galactic plane. 
Electron 
densities as a function of $R$ are presented in TC93 (in their Fig.~3).

The fifth component becomes dominant up to a Galactocentric distance of 0.6 kpc. 
In this area, numerous point sources of ionization are responsible for the 
total Galactic center (GC) component contribution, especially within the 
nuclear bulge ($r \leq 250$ pc). 

At a dynamical center of the Galaxy, SgrA$^*$ -- a nonthermal synchrotron 
radio-source coinciding with the estimated $2.6 \times 10^6 M_{\odot}$ black 
hole [6], lies surrounded with thermal plasma -- SgrA West, 
ionized by a cluster of few tens of young, hot ($T_{\rm eff}=3.5 \times 10^4$  K), 
luminous O stars within a central ($r=1$ pc) cavity [14]. SgrA West consists of
 a minispiral and an extended component with
 $n_e^{\rm sp} \sim 10^4 \; {\rm cm}^{-3}$ and $n_e^{\rm ext} \sim 10^3 \; {\rm cm}^{-3}$ 
[15]. Some 
local features of ionized gas such as the Bullet (4" northwest of SgrA$^*$) 
[16] and the Sickle [17] contribute to the total 
electron density  near GC with $3 \times 10^4 \; {\rm cm}^{-3}$ and 150 ${\rm 
cm}^{-3}$ respectively. 

At a distance of 30 pc behind the GC, lays SgrA East -- the probable supernova 
remnant colliding with a molecular cloud at 200 pc from GC, contributing with 
$n_e=6\; {\rm cm}^{-3}$ electron density.
It could be heated by O stars formed as a result of 
collision~[19].

Another large ionized area near GC is a Sgr B2 -- the largest luminous H 
II/mol\-ec\-ul\-ar cloud region at 100 pc from GC surrounding 
the circumnuclear disc (2-12 pc from GC) in which the SgrA West is embedded. 
SgrB2 could be interpreted as an extended accretion disk around a central 
dormant black hole [20].

If we now wish to study the ionization in outer parts of the Galaxy we
 will encounter the lack of star forming regions in the outskirts of spiral 
galaxies [21]. From the radio observations 
(21 cm) it was noticed that HI disks in at least two galaxies (M33 and NGC3198)
 have sharp edges [1]. 
HI column density drops from few times 
$10^{19}$ cm$^{-2}$ to few times  $10^{18}$ cm$^{-2}$ within 1 to 2 kpc.
 When the incident photon flux $F$ is monochromatic and the ionized gas is smoothly distributed 
 (with density $n$ and thickness $b$) the edge will occur 
when $N\sim {F\over \alpha n}$ where $\alpha$ is the recombination coefficient.
If one wishes to estimate the intergalactic hydrogen-ionizing 
flux $F$, one can  measure H$\alpha$ surface brightness of clouds opaque to
 $F$ exposed to this flux -- each hydrogen recombination produces 0.46 
H$\alpha$ photons on average (e.g. [1]). After extensive studies 
an upper limit for ionizing flux is currently thought to  be:
$F_{ext}\le 6\times 10^5\; {\rm cm}^{-2}\,{\rm s}^{-1}$
[1].
One can generally estimate $F$ by measuring the recombination emission induced 
by it [22, 23]. 
Although uncertainties connected with this quantity are large, their
relevance is largely beyond the scope of this paper, since metagalactic flux
is the dominant ionization source only at large heights above the plane
of the disk and at large galactocentric radii, i.e. in the halo gas. 
We briefly mention that an interesting candidate for ionizing source present in all galaxies
could be the massive decaying neutrino, which not only solves 
the problem of the ionization, but also accounts for the dark mass present in
every spiral galaxy.
There exist a large number of alternative sources of intergalactic ionization: 
population III stars, elliptical galaxies, dwarf galaxies, 
starbursts, quasars concealed by dust obscuration in
 intervening galaxies, reflection-dominated hard x-ray sources, accreting 
black holes and supernova-driven winds from early galaxies ([1],
 and references therein), or non-radiative sources like large-scale 
intergalactic shocks. Most of them, fortunately, are not operational in
late epochs of galactic evolution, and their respective influences will be 
considered in the extension of the present work. 

The significance of these results lies mainly in exciting possibility of 
direct testing of models through high-sensitivity observations of 
recombination lines [24]. One of
problems facing any such observations, especially in cases of small 
inclination angles, is strong signal originating in the fluorescing haloes
interfering with light from the disk. We roughly compare 
two intensities using the estimate of local ($z=0$) intensity of H$\alpha$ 
halo emission from [23], and form:
\vskip-0.5cm
$$
\frac{I_{\alpha}^{\rm disk}}{I_{\alpha}^{\rm halo}} \approx 42.9 {\cal I} 
\left( \frac{N_{\rm HI}}{10^{15} \; {\rm cm}^{-2}} \right)^{-1} T_4^{-0.73},
\eqno(2)$$
\vskip-0.15cm
\noindent where ${\cal I}$ is the value  of the numerically solved integral
 $\int n_e^2 d s=2.75 \, T_4^{0.9}I_{\alpha} \; {\rm cm^{-6}\,pc}$ [26] and expression
  for $n_e$ 
is taken from the rhs of the eq.~(1) using first two terms. 
$N_{\rm HI}$ is the neutral hydrogen column density along the line of 
sight {\it through the halo only}.
In the Figure 1. we plot the dependence of $I_\alpha$, that
is interstellar H$\alpha$ intensity in rayleighs (1 R=$10^6/4\pi$ 
 photons cm$^{-2}$ s$^{-1}$ sr$^{-1}$), on the Galactocentric distance, $r$.
 Although poorly known, $N_{\rm HI}$ is
expected to be $\sim 10^{15}$ cm$^{-2}$, characteristic of the average 
Ly$\alpha$ absorption originating in haloes of normal galaxies
at low redshift [25]. Therefore, the signal
from the disk should be strong enough to be analyzed in most of the typical 
cases. Of course, the kinematic properties of the emission lines 
originating in halo and disk will also  be different, helping 
practical discrimination between the two.

We note that Reynolds et al. [27] obtain for $I_\alpha$ at $|z| \approx 1$ kpc
the following value: $4\pi J \approx 2\times 10^6$ photons cm$^{-2}$ s$^{-1}$, where $J$ 
is the incident Lyman continuum flux,
that can be  compared with our estimate depicted in the Figure 1. 
The project that is currently operating -- Wisconsin H$\alpha$ Mapper (WHAM)
[28] will provide the data on the diffuse ionized hydrogen through the 
optical H$\alpha$ line. In a subsequent work, we intend to include all
 terms in the eq. (1), in order to get complete picture of expected H$\alpha$ emission,
  by solving the integral $\int n_e^2ds$.  

\smallskip
\noindent{\bf ACKNOWLEDGEMENT:} We acknowledge the help of Prof. James Cordes 
in obtaining the code written for the paper TC93.

\smallskip
\smallskip
\noindent {\bf REFERENCES}
\smallskip

\item{}\kern-\parindent{[1] D.W. Sciama, 
{\it Modern Cosmology and the Dark Matter Problem}, CUP, Cambridge (1993).} 

\item{}\kern-\parindent{[2] D.W. Sciama, {\it Astrophys. J.} {\bf 488} (1997) 234.}

\item{}\kern-\parindent{[3] P. Maloney, {\it Astrophys. J.} {\bf 414} (1993) 41.}
 
\item{}\kern-\parindent{[4] R.J. Reynolds and D.P. Cox, {\it Astrophys. J. Lett.}
 {\bf 400} (1992) L33.} 

\item{}\kern-\parindent{[5] Y. Sofue, {\it Astrophys. J.} {\bf 458} (1996) 120.}

\item{}\kern-\parindent{[6] G.C. Bower and D.C. Backer, 
{\it Astrophys. J.}~{\bf 496}~(1998)~97.}

\item{}\kern-\parindent{[7] H. Falcke, W.M. Goss, H. Matsuo,
 P. Teuben, J.H. Zhao and R.  Zylka,
{\it Astrophys. J.} {\bf 499}~(1998)~731.}

\item{}\kern-\parindent{[8] J.H. Taylor and J.M. Cordes, {\it Astrophys. J.} {\bf 
411} (1993) 674 (TC93).}

\item{}\kern-\parindent{[9] T.J.W. Lazio and J.M. Cordes, {\it Astrophys. J.} {\bf 
497} (1998) 238 (LC98).}

\item{}\kern-\parindent{[10] F. Combes, P. Boiss\'e, A. Mazure and A. Blanchard, 
{\it Galaxies and 
Cosmology}, Springer Verlag, Berlin Heidelberg New York (1995).}

\item{}\kern-\parindent{[11] T.E. Nordgren, J.M. Cordes and Y. Terzian, {\it Astronom. J.}
 {\bf 104} (1992) 1465.} 

\item{}\kern-\parindent{[12] R.J. Reynolds, {\it Astrophys. J. Lett.} 
{\bf 349} (1990) L17.} 

\item{}\kern-\parindent{[13] R.J. Reynolds, {\it Astrophys. J. Lett.}
 {\bf 392} (1992) L35.}

\item{}\kern-\parindent{[14] R. Zylka, P.G. Mezger, D. Ward-Thompson, W. J. Duschl 
and H. Lesch, {\it A \& A} {\bf 297} (1995) 83.}

\item{}\kern-\parindent{[15] T. Beckert, W. J. Duschl, P.G. Mezger and R. Zylka, 
{\it A \& A} {\bf 307} (1996) 450.}

\item{}\kern-\parindent{[16] F. Yusef-Zadeh, D. A. Roberts and J. Biretta, {\it 
Astrophys. J.} {\bf 499}~(1998)~159.}

\item{}\kern-\parindent{[17] F. Yusef-Zadeh, D. A. Roberts and M. Wardle, {\it 
Astrophys. J.} {\bf 490}~(1997)~83.}

\item{}\kern-\parindent{[18] F. Yusef-Zadeh, 
W. Purcell and E. Gotthelf, {\it IAUS} {\bf 184}~(1997)~196.}                                                 

\item{}\kern-\parindent{[19] M. Wardle, F. Yusef-Zadeh and T. R. Geballe, {\it 
Astrophys. J. Lett.} submitted, preprint astro-ph/9804146 (1998).}

\item{}\kern-\parindent{[20] M. A. Gordon, U. Berkermann, P. G. Mezger, R. Zylka, 
C. G. T. Haslam, E. Kreysa, A. Sievers and R. Lemke, {\it A \& A} {\bf 280} 
(1993) 208.}

\item{}\kern-\parindent{[21] E. Corbelli and E.E. Salpeter, {\it Astrophys. J.} 
{\bf 419} (1993) 104.}

 \item{}\kern-\parindent{[22] C.J. Hogan and R.J. Weymann, 
 {\it MNRAS}, {\bf 225}~(1987)~1p.}

\item{}\kern-\parindent{[23] M. M. \'Cirkovi\'c, J. Bland-Hawthorn and S. 
Samurovi\'c, {\it MNRAS}, (1998) submitted.}

\item{}\kern-\parindent{[24] J. Bland-Hawthorn, K. C. Freeman and P. J. Quinn
{\it Astrophys. J.} {\bf 490} (1997) 143.}

 \item{}\kern-\parindent{[25] H.W. Chen, K.M. Lanzetta, J.K. Webb and X. 
Barcons {\it Astrophys. J.} {\bf 498}~(1998)~77.}

 \item{}\kern-\parindent{[26] R.J. Reynolds {\it Astrophys. J.} {\bf 372} (1991) L17.}

\item{}\kern-\parindent{[27] R.J. Reynolds, S.L. Tufte, D.T. Kung, P.R.
 McCullough and C. Heiles
{\it Astrophys J.}  {\bf 448}~(1995)~715.}

\item{}\kern-\parindent{[28] R.J. Reynolds, S.L. Tufte, L.M. Haffner, 
K. Jaehnig and J.W. Percival:
{\it PASA}, {\bf 15}~1998~14.}

\vfill\eject

\hskip0.1cm
\vskip-0.5truecm
\centerline{\psfig{file=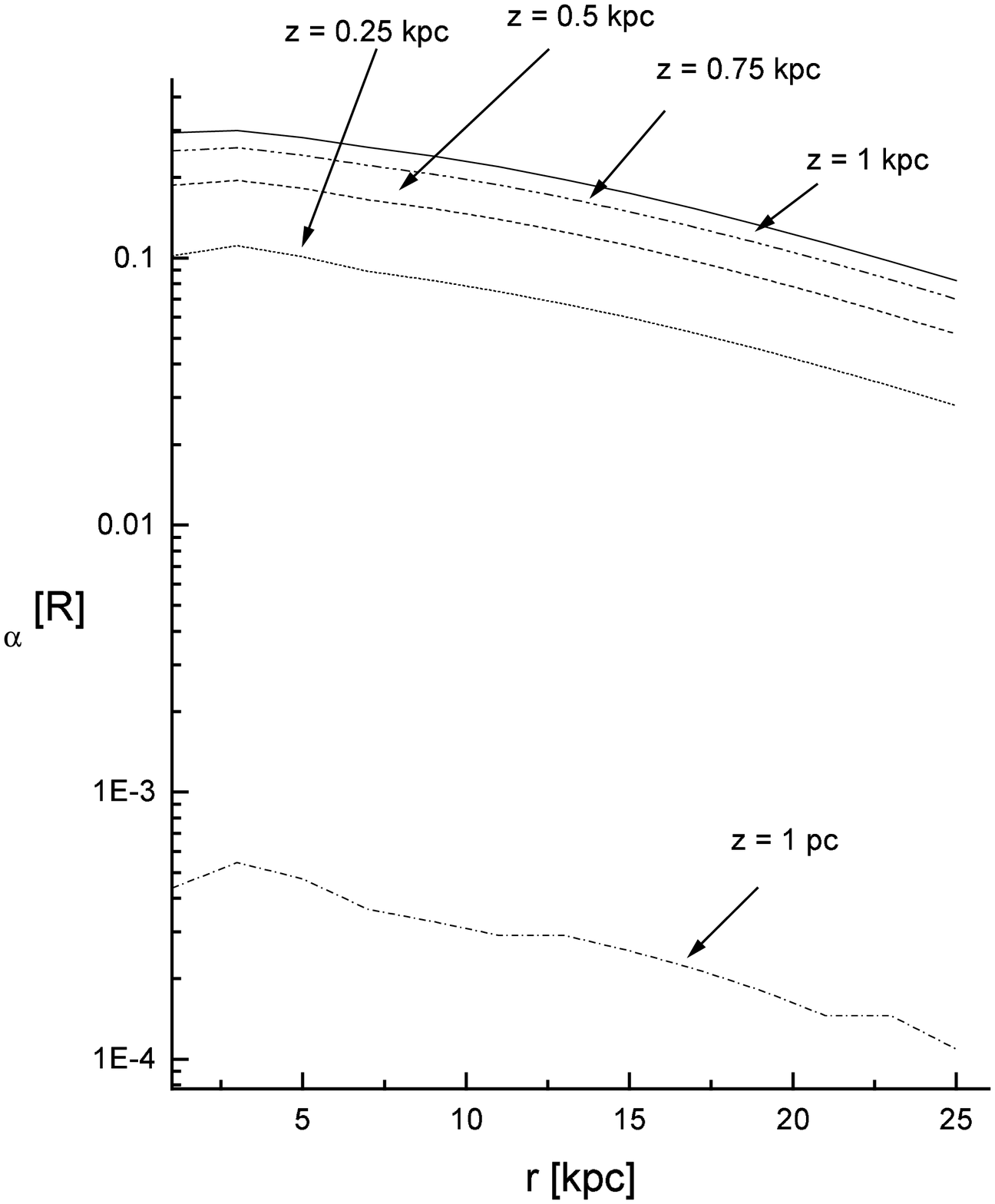,height=12truecm}}

\bigskip
\bigskip
\bigskip
\noindent  Figure 1.
  The interstellar H${}{\seveni \alpha}$ intensity I${\seveni {}_\alpha}$
in rayleighs  is plotted 
against the Galactocentric distance r in kiloparsecs. Different 
values for the height above the Galactic plane,  z
 (where z goes from
the disk to 1 kpc) are taken into account.
                                                                     
\medskip
\bye